\documentclass{article}
\usepackage{spconf,amsmath,graphicx,amssymb}
\usepackage{color,multirow}
\usepackage{cite,titlesec}

\titleformat{\subsubsection}[runin]{}{}{}{}[]

\title{D3Net: Densely connected multidilated DenseNet for Music source separation}
\name{
Naoya Takahashi, Yuki Mitsufuji
}

\address{Sony Corporation, Japan}
\begin{document}
\ninept
\fontsize{9.5pt}{12.6pt}\selectfont

\maketitle
\begin{abstract}
Music source separation involves a large input field to model a long-term dependence of an audio signal. Previous convolutional neural network (CNN) -based approaches address the large input field modeling using sequentially down- and up-sampling feature maps or dilated convolution. In this paper, we claim the importance of a rapid growth of a receptive field and a simultaneous modeling of multi-resolution data in a single convolution layer, and propose a novel CNN architecture called densely connected dilated DenseNet (D3Net). D3Net involves a novel multi-dilated convolution that has different dilation factors in a single layer to model different resolutions simultaneously. By combining the multi-dilated convolution with DenseNet architecture, D3Net avoids the aliasing problem that exists when we naively incorporate the dilated convolution in DenseNet. 
Experimental results on MUSDB18 dataset show that D3Net achieves state-of-the-art performance with an average signal to distortion ratio (SDR) of 6.01 dB.\footnote{The conference version of this paper, which includes extended works, is available. Please refer to Naoya Takahashi et al. "Densely connected multidilated convolutional networks for dense prediction tasks", CVPR2021}
\end{abstract}

\begin{figure*}[t]
  \centering
  \includegraphics[width=\linewidth]{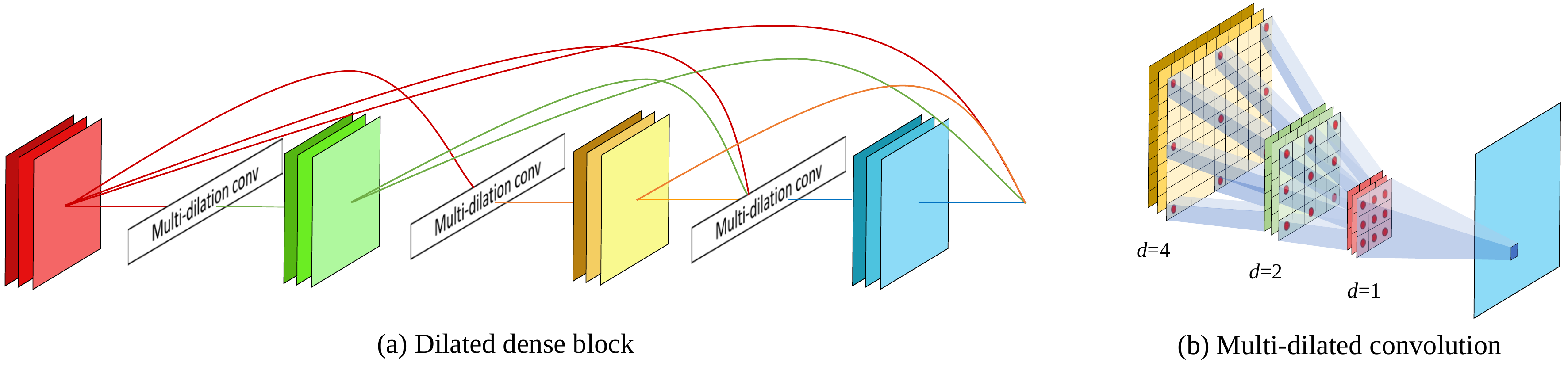}
  \caption{Illustration of D2 block. (a) The connectivity pattern is the same as in DenseNet except that the D2 block involves the multi-dilated convolution. (b) Illustration of the multi-dilated convolution at the third layer. To produce a single feature map, it involves multiple dilation factors depending on the input channel. For clarity, we omit the normalization and nonlinearity from the illustration.
}
  \label{fig:d2block}
\end{figure*}

\section{Introduction}
\label{sec:intro}
Music source separation (MSS) has been intensively studied and neural-network-based approaches have shown impressive progress in recent years. Many types of neural network architecture have been proposed, including feedforward fully connected networks (FNNs)\cite{Nugraha15, Uhlich15},  recurrent neural networks (RNNs) \cite{Uhlich17}, convolutional neural networks (CNNs) \cite{Takahashi17,Stoller18WavU,Takahashi18PhaseNet,defossez2019demucs,Samuel20,spleeter2019} and their combinations \cite{Takahashi18MMDenseLSTM,Liu19}.
In particular, CNNs have attracted great attention because of their superior performance, parameter efficiency and generality for different types of data. Takahashi et.~al.~\cite{Takahashi18MMDenseLSTM} applied a CNN architecture with a dense skip connectivity pattern, called DenseNet \cite{Huang17Densenet}, to MSS and obtained state-of-the-art results in SiSEC 2018 \cite{sisec2018}.  Such dense connectivity allows maximum information flow and deeper CNN while keeping the model size small by efficiently reusing intermediate representations of preceding layers.


One of the benefits of a deeper CNN is its larger receptive field that allows a large context to be modeled, which is important since audio signals can have long time and wide frequency band dependences. 
Although the receptive field grows linearly with the number of layers stacked, it is not the optimal way to increase the receptive field only by stacking convolutional layers, as it requires too many layers to cover a sufficiently large input field to model global information, making the network training too difficult. 
%
A popular approach to incorporating a large context is to repeatedly downsample intermediate network outputs and apply operations in lower resolution representations. The low-resolution representations are again upsampled to recover the lost resolution while carrying over the global perspective from downsampled layers \cite{Takahashi17, Stoller18WavU, defossez2019demucs, Takahashi18MMDenseLSTM}. 
Another approach is a dilated convolution, which is shown to be effective for audio generation and MSS tasks \cite{Aaron2016WN, defossez2019demucs,Liu19}. The dilation factors are set to grow exponentially with the number of layers stacked and, therefore, the networks cover the large receptive field with a small number of layers.
Although the down--upsampling structure and dilated convolution allow a large receptive field, each layer in the network sees only one resolution at a time. However, the simultaneous consideration of  the local and global information can be useful, e.g., the local structure can be more precisely estimated by using global structure information and vise versa.
DenseNet partially addresses this problem by means of the dense skip connectivity that allows the direct aggregation of features from early layers and features in later layers within a single convolution layer.
However, it may still be too slow to transform local features to global features and it is inefficient to have many parameters, especially for high-resolution data.

In this work, we combine the advantages of DenseNet and dilated convolution, and propose a novel network architecture called dilated DenseNet (D2Net). To properly combine DenseNet with the dilated convolution, we propose a multidilated convolution layer that has a multiple dilation factor within a single layer. The dilation factor depends on which skip connection the channels come from, as shown in Fig.\ref{fig:d2block}. The multidilated convolution can prevent the aliasing that occurs when a standard dilated convolution is applied to feature maps with receptive fields smaller than the dilation factor. Although a naive combination of DenseNet with dilation has already been proposed \cite{Fuchs19}, standard dilated convolutions are used and dilation factors are determined depends on the layer depth, which causes considerable aliasing. In contrast, we show the effectiveness of the proposed multidilated convolution in our ablation study.
Furthermore, we propose a nested architecture of dilated dense blocks to effectively repeat dilation factors multiple times with dense connections that ensure the sufficient depth required for modeling each resolution. 
We call the nested architecture densely connected dilated DenseNet (D3Net).


The contributions of this work are summarized below:
{
\setlength{\leftmargini}{20pt} 
\begin{enumerate}
\item We claim that a naive incorporation of dilation in DenseNet architecture can cause a significant aliasing problem, and propose a multidilated convolution layer to properly incorporate the dilated convolution into DenseNet. 
\item We further introduce the D3Net architecture of nested dilated dense blocks to effectively apply different dilation factors multiple times. 
\item We experimentally show the effectiveness of the proposed architectures. The D3Net achieves state-of-the-art results on MUSDB18 dataset. \end{enumerate}
}

\section{Multidilated convolution for DenseNet}
\label{sec:d2}
\begin{figure}[t]
  \centering
  \includegraphics[width=\linewidth]{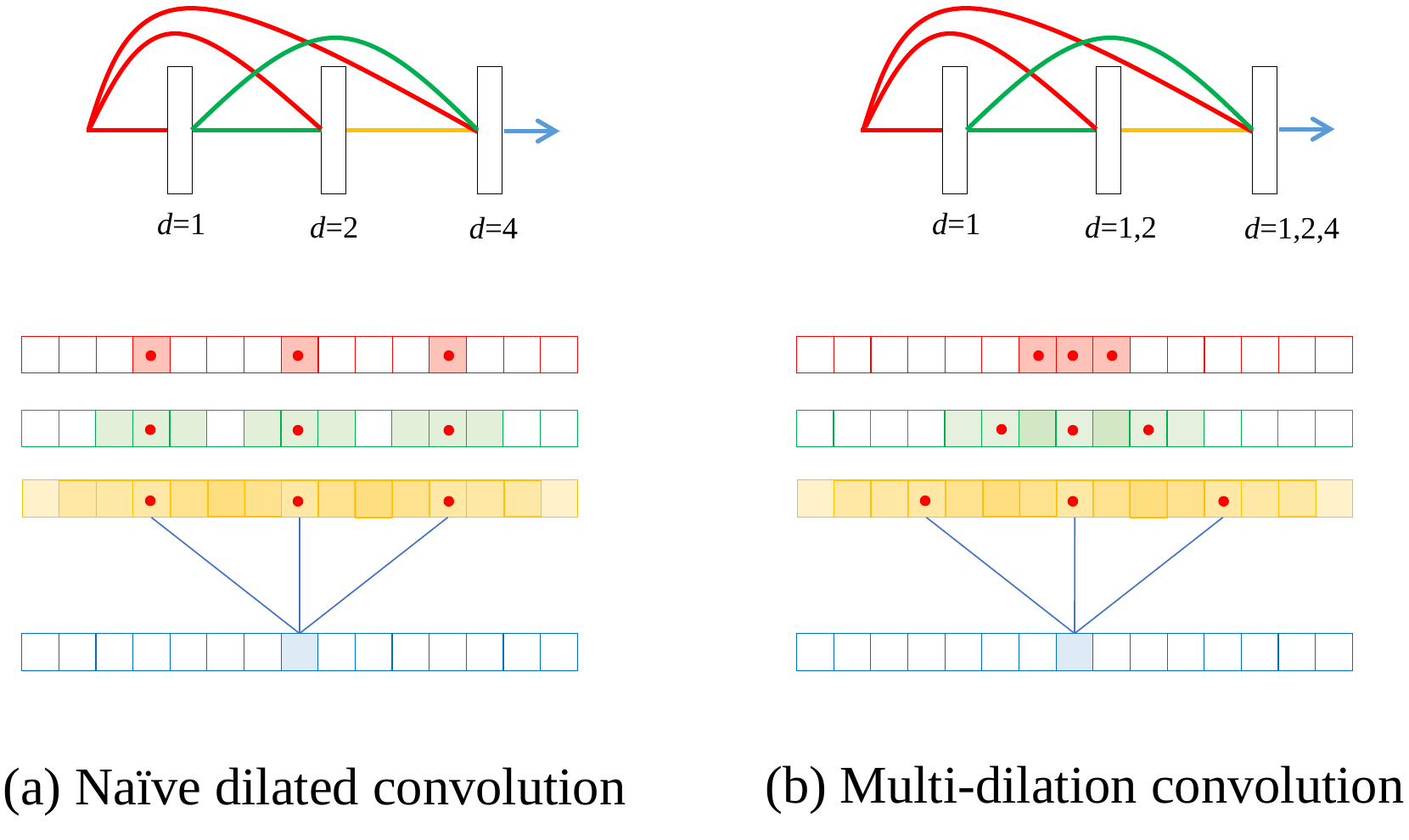}
  \caption{Visualization of receptive fields at the third layer of (a) naive integration of dilated convolution and (b) proposed multi-dilated convolution (in the case of one dimension). Red dots denote the points to which filters are applied, and the colored background shows the receptive field covered by the red dot.}
  \label{fig:d2rf}
\end{figure}

In DenseNet, the outputs of the $l$th convolutional layer $x_l$ are computed using filters $k_l$ and outputs of all preceding layers as
\begin{equation}
x_l = \psi([x_0, x_1, \cdots, x_{l-1}]) \circledast k_l,
\end{equation}
where $\psi()$ denotes the composite operation of batch normalization and nonlinearity, $[x_0, x_1, \cdots, x_{l-1}]$ the concatenation of feature maps from $1, \cdots, l-1$ layers, and $\circledast$ the convolution.
A naive way of incorporating dilated convolution is to replace the convolution $\circledast$ with the dilated convolution $\circledast_d$ with the dilation factor $d=2^{l-1}$. However, this causes a severe aliasing problem; for instance, at the third layer, input is subsampled with 4 sample intervals without any anti-aliasing filtering because of the skip connections. Assuming that the kernel size is 3, only the path that passes through all convolution operations without any skip connection covers the input field without omission and all other paths from skip connections have \textit{blind spots} in their receptive fields that inherently make it impossible for proper ant-aliasing filters to be learned in the preceding layers (Fig.~\ref{fig:d2rf}a). 
To overcome this problem, we propose the multidilated convolution $\circledast^m_l$ defined as
\begin{equation}
Y_l \circledast^m_l k_l = \sum_{i=0}^{l-1} y_i \circledast_{d_i} k^i_l,
\end{equation}
where $Y_l = [y_0, y_1, \cdots, y_{l-1}] = \psi([x_0, x_1, \cdots, x_{l-1}])$ is the composite layer output,  $k^i_l$ the subset of filters that correspond to the $i$th skip connection, and $d_i = 2^i$. As depicted in Fig.~\ref{fig:d2rf}b, DenseNet with the proposed multidilated convolution has different dilation factors depending on which layer the channel comes from. This allows the receptive field to cover the input field without the loss of coverage between the samples the filters to be applied and, hence, to learn proper filters to prevent aliasing. 
One advantage of the dilated dense block (D2 block) is its ability to integrate information from very local to exponentially large receptive field within a single layer. This fast information flow provides more flexibility in modeling information in a wide range of resolutions.

Note that the multidilation convolution is not equivalent to applying the multibranch convolution where convolutions with different dilation factors are applied to the same input feature maps, similar to the Inception block \cite{McMahan18, Szegedy15, Szegedy16}, again causing the aliasing problem.

\section{D3Net}
\label{sec:d3}
\begin{figure}[t]
  \centering
  \includegraphics[width=\linewidth]{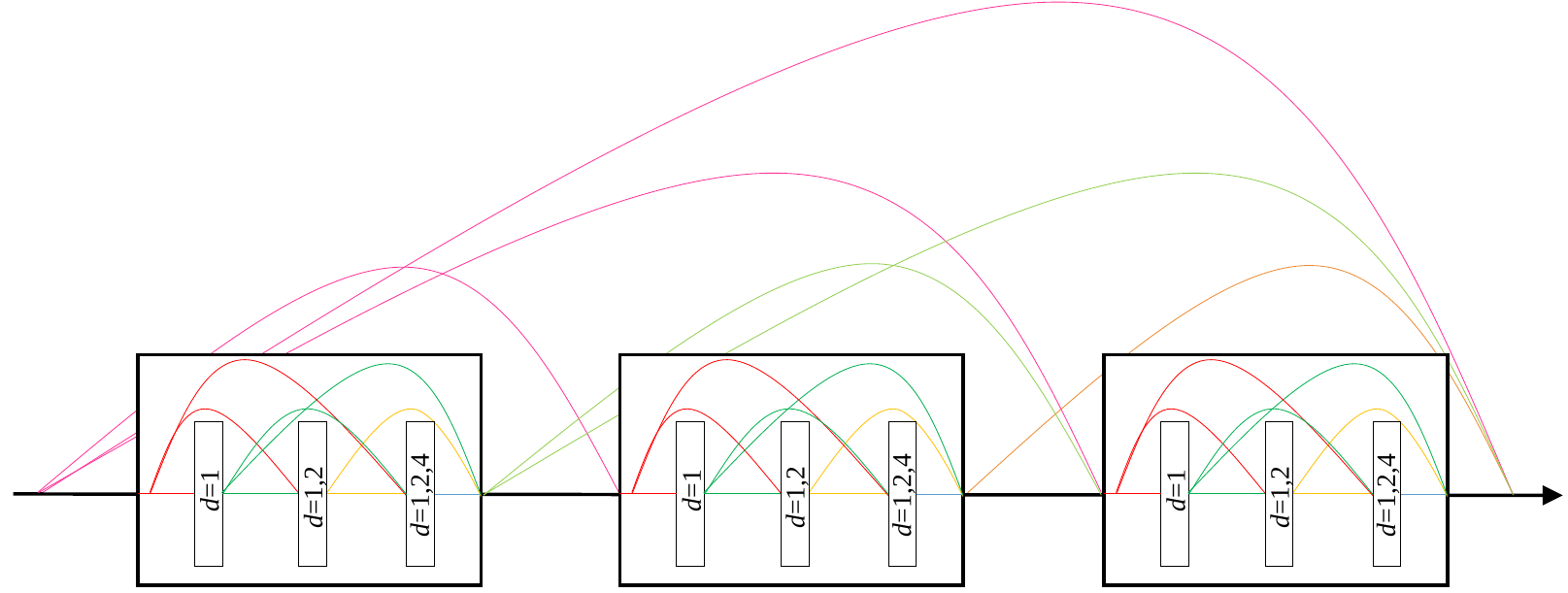}
  \caption{D3 block densely connects D2 blocks with repeated dilation pattern.}
  \label{fig:D3 block}
\end{figure}

Although the D2 block provides an exponentially large receptive field as the number of layers increases, it is also worthwhile to provide sufficient flexibility to transform feature maps in each resolution. In WaveNet \cite{Aaron2016WN}, dilation factors are reset to one after several layers are stacked and repeated; that is, the dilation factor in the $l$th layer is given by $d_l=2^{l-1~mod~M}$, where $mod$ is the modulo operation and $M$ is the number of layers at which the dilation factor is doubled. Inspired by this work, we propose a nested architecture of D2 blocks as shown in Fig.~\ref{fig:D3 block}. D2 blocks are considered as single composite layers and are densely connected in the same way as within the D2 block itself. We also employ a channel reduction mechanism at the end of each D2 block to mitigate the growth of an excessive number of channels and thus improve computational efficiency. The channel reduction can be performed by either a $1\times1$ convolution or simply passing the output of the last $N$ layers' to the next block. In this work, we take the latter approach since performance characteristics of both methods are similar, but the former approach requires slightly more computations. Note that without the channel reduction, the architecture is reduced to a standard dense connection with repeated multidilation factors.

\section{Experiments}
\textbf{Dataset} \hspace{1mm}
We evaluated the proposed method using the MUSDB18 dataset, prepared for SiSEC 2018 \cite{sisec2018}. 
In the dataset, approximately 10 hours of professionally recorded 150 songs in stereo format at 44.1kHz are available. For each song, a mixture and its four sources, {\it bass, drums, other} and {\it vocals}, are provided and thus, the task is to separate the four sources from the mixture. We adopted the official split of 100 and 50 songs for {\it Dev} and {\it Test} set, respectively.
Short-time Fourier transform (STFT) magnitude frames of the mixture, windowed at 4096 samples with 75\% overlap, with data augmentation \cite{Uhlich17} were used as inputs. 
\vspace{3mm}\\
\textbf{Training} \hspace{1mm}
The four networks for each source instrument were trained to estimate the source spectrogram by minimizing the mean square error with the Adam optimizer for 50 epochs. The patch length was set to 256 frames; thus, the dimensions of input were $2\times256\times2049$. The batch size was set to 6. The learning rate was initially set to 0.001 and annealed to 0.0001 at 40 epochs.
\vspace{3mm}\\
\textbf{Model architecture} \hspace{1mm}
\begin{table*}[t]
    \caption{\label{tab:d3arch} Proposed architectures. All D3 blocks have 3$\times$3 kernels with growth rate $k$, $L$ layers, and $M$ D2 blocks.}
    \vspace{2mm}
    \centerline{
      \footnotesize
      \tabcolsep=3px
      \begin{tabular}{ c | c | c | c | c | c | c | c | c | c | c} 
        \hline
        \multirow{2}{*}{Layer} & \multirow{2}{*}{scale} & \multicolumn{3}{c|}{Vocals, Other} & \multicolumn{3}{c|}{Drums} & \multicolumn{3}{c}{Bass} \\
        \cline{3-11}
        & & low & high & full & low & high & full & low & high & full \\
        \hline
        band split index & \multirow{3}{*}{1} & 1-256 & 257-1600 &  - & 1-128 & 128-1600 &  - & 1-192 & 192-1600 &  -\\
        conv (t$\times$f,ch) &  & 3$\times$3, 32 & 3$\times$3, 8  & 3$\times$3, 32 & 3$\times$3, 32 & 3$\times$3, 8  & 3$\times$3, 32 & 3$\times$3, 32 & 3$\times$3, 8  & 3$\times$3, 32 \\
        D3 block 1 (k,L,M) & & 16, 5, 2 & 2, 1, 1 & 13, 4, 2 & 16, 5, 2 & 2, 1, 1 & 13, 4, 2 & 16, 5, 2 & 2, 1, 1 & 10, 4, 2\\
        \hline                 
        down sample  & \multirow{2}{*}{$\frac{1}{2}$} &  \multicolumn{3}{c|}{avg. pool $2\times2$} & \multicolumn{3}{c|}{avg. pool $2\times2$} & \multicolumn{3}{c}{avg. pool $2\times2$}\\
        \cline{3-11}
        D3 block 2 (k,L,M) & & 18, 5, 2 & 2, 1, 1 & 14, 5, 2  & 18, 5, 2 & 2, 1, 1 & 14, 5, 2 & 18, 5, 2 & 2, 1, 1 & 10, 5, 2\\
        \hline
        down sample  & \multirow{2}{*}{$\frac{1}{4}$} &  \multicolumn{3}{c|}{avg. pool $2\times2$} & \multicolumn{3}{c|}{avg. pool $2\times2$} & \multicolumn{3}{c}{avg. pool $2\times2$}\\
        \cline{3-11}
        D3 block 3 (k,L,M) & & 20, 5, 2 & 2, 1, 1 & 15, 6, 2 &  20, 5, 2 & 2, 1, 1 & 15, 6, 2  & 18, 5, 2 & 2, 1, 1 & 12, 6, 2 \\
        \hline
        down sample  & \multirow{2}{*}{$\frac{1}{8}$} &  \multicolumn{3}{c|}{avg. pool $2\times2$} & \multicolumn{3}{c|}{avg. pool $2\times2$}  & \multicolumn{3}{c}{avg. pool $2\times2$}\\
        \cline{3-11}
        D3 block 4 (k,L,M) & & 22, 5, 2 & 2, 1, 1 & 16, 7, 2 & 22, 4, 2 & 2, 1, 1 & 16, 7, 2& 20, 5, 2 & 2, 1, 1 & 14, 7, 2 \\
        \hline
        down sample  & \multirow{2}{*}{$\frac{1}{16}$} &  \multicolumn{3}{c|}{avg. pool $2\times2$} & \multicolumn{3}{c|}{avg. pool $2\times2$} &  \multicolumn{3}{c}{avg. pool $2\times2$} \\
        \cline{3-11}
        D3 block 5 (k,L,M) & & - & - & 17, 8, 2 &  - & - & 16, 8, 2 & - & - & 16, 8, 2 \\
        \hline
        up sample  & \multirow{3}{*}{$\frac{1}{8}$} & \multicolumn{3}{c|}{t.conv $2\times2$} &  \multicolumn{3}{c|}{t.conv $2\times2$} & \multicolumn{3}{c}{t.conv $2\times2$} \\
        \cline{3-11}
        concat. & & - & - & D3 block 4 & - & - & D3 block 4 & - & - & D3 block 4 \\
        D3 block 6 (k,L,M) & & -& - & 16, 6, 2& -& - & 16, 6, 2& -& - & 14, 6, 2 \\
		\hline
        up sample  & \multirow{3}{*}{$\frac{1}{4}$} & \multicolumn{3}{c|}{t.conv $2\times2$} &  \multicolumn{3}{c|}{t.conv $2\times2$} & \multicolumn{3}{c}{t.conv $2\times2$} \\
        \cline{3-11}
        concat. & & D3 block 3 & D3 block 3 & D3 block 3 & D3 block 3 & D3 block 3 & D3 block 3 & D3 block 3 & D3 block 3 & D3 block 3\\
        D3 block 7 (k,L,M) & & 20, 4, 2& 2, 1, 1 & 14, 5, 2 & 20, 4, 2& 2, 1, 1 & 14, 6, 2 & 18, 4, 2& 2, 1, 1 & 12, 6, 2\\
		\hline
        up sample  & \multirow{3}{*}{$\frac{1}{2}$} & \multicolumn{3}{c|}{t.conv $2\times2$} & \multicolumn{3}{c|}{t.conv $2\times2$}  & \multicolumn{3}{c}{t.conv $2\times2$}\\
        \cline{3-11}
        concat. & & D3 block 2 & D3 block 2 & D3 block 2 & D3 block 2 & D3 block 2 & D3 block 2 & D3 block 2 & D3 block 2 & D3 block 2 \\
        D3 block 8 (k,L,M) & & 18, 4, 2 & 2, 1, 1 & 12, 4, 2  & 18, 4, 2 & 2, 1, 1 & 12, 4, 2 & 16, 4, 2 & 2, 1, 1 & 8, 4, 2\\
		\hline
        up sample  & \multirow{3}{*}{1} & \multicolumn{3}{c|}{t.conv $2\times2$} &  \multicolumn{3}{c|}{t.conv $2\times2$}  & \multicolumn{3}{c}{t.conv $2\times2$} \\
        \cline{3-11}
        concat. & & D3 block 1 & D3 block 1 & D3 block 1 & D3 block 1 & D3 block 1 & D3 block 1 & D3 block 1 & D3 block 1 & D3 block 1\\
        D3 block 9 (k,L,M) & & 16, 4, 2 & 2, 1, 1 & 11, 4, 2  & 16, 4, 2 & 2, 1, 1 & 11, 4, 2 & 16, 4, 2 & 2, 1, 1 & 8, 4, 2\\        
        \hline
        concat. (axis) & \multirow{4}{*}{1} & \multicolumn{2}{c|}{freq} & - & \multicolumn{2}{c|}{freq} & - & \multicolumn{2}{c|}{freq} & - \\
        \cline{3-11}
        concat. (axis) &  & \multicolumn{3}{c|}{channel} &  \multicolumn{3}{c|}{channel} & \multicolumn{3}{c}{channel}\\
        \cline{3-11}
        d2 block (k,L) & & \multicolumn{3}{c|}{12, 3} & \multicolumn{3}{c|}{12, 3} & \multicolumn{3}{c}{12, 3} \\   
        \cline{3-11}
        gate conv (t$\times$f,ch) &   & \multicolumn{3}{c|}{$3\times3$, 2} &  \multicolumn{3}{c|}{$3\times3$, 2}&  \multicolumn{3}{c}{$3\times3$, 2}\\
        \hline
      \end{tabular}
    }
\end{table*}
Following \cite{Takahashi17,Takahashi18MMDenseLSTM}, in which the best results obtained in SiSEC 2018 were reported, we used the multiscale multiband architecture in which band-dedicated modules and a full band module, each with a bottleneck encoder--decoder architecture with skip connections, are placed. The network configuration is shown in Table \ref{tab:d3arch}. The network outputs are used to calculate the multichannel Wiener filter (MWF) to obtain the final separations, as commonly performed in frequency domain audio source separation methods \cite{Takahashi18MMDenseLSTM, Uhlich17, Liu19, Samuel20}.
\begin{table}[t]
\caption{\label{tab:mss} {\it SDR values for MUSDB18 dataset. SDR values are median of median SDR of each song. '*' denotes method operating in time domain.}}
\vspace{2mm}
\centering{
\resizebox{\linewidth}{!}{
\begin{tabular}{ c | c c c c c c} 
\hline
\multicolumn{1}{c|}{} & \multicolumn{6}{c}{SDR in dB}\\
Method      &	Vocals	&	Drums	& Bass & Other & Acco. & Avg.\\
\hline\hline
TAK1 {\scriptsize (MMDenseLSTM)} \cite{Takahashi18MMDenseLSTM}\ 	&	6.60 & 6.43 & 5.16 & 4.15 & 12.83 &5.59\\
UHL2 {\scriptsize(BLSTM ensemble)} \cite{Uhlich17}\ 	&	5.93 & 5.92 & 5.03 & 4.19 & 12.23 &5.27\\
GRU dilation 1 \cite{Liu19}\ 	& 6.85 & 5.86 & 4.86 & \textbf{4.65} & 13.40 & 5.56\\
UMX \cite{stoter19}\ 	& 6.32 & 5.73 & 5.23 & 4.02 & - & 5.33\\
demucs* \cite{defossez2019demucs}\ 	& 6.29 & 6.08 & 5.83 & 4.12 & - & 5.58\\
Meta-TasNet* \cite{Samuel20} 	& 6.40 & 5.91 & 5.58 & 4.19 & - & 5.52\\
Nachmani et. al.* \cite{Nachmani20} & 6.92 & 6.15 &  \textbf{5.88} & 4.32 & - & 5.82\\
\hline
D3Net w/o dilation      & 6.86 & 6.37	& 4.97 & 4.21 & 13.19 & 5.60\\
D3Net standard dilation	& 7.12	& 6.61 & 5.19 & 4.53  & 13.39 & 5.86\\
{\bf D3Net} (proposed)	& \textbf{7.24}	& {\bf 7.01} & 5.25 & 4.53 & {\bf 13.52} & \textbf{6.01}\\
\hline
\end{tabular}
}
}
\end{table} 
\vspace{3mm}\\
\textbf{Results} \hspace{1mm}
The signal-to-distortion ratio (SDR) of our proposed method and existing state-of-the-arts methods are compared in Table \ref{tab:mss}. The SDRs were computed using  the {\it museval} package \cite{sisec2018} and median SDRs are reported. 
TAK1 \cite{Takahashi18MMDenseLSTM} and UHL2 \cite{Uhlich17} are the two best performing methods in SiSEC 2018 (among submissions that do not use external data) and the network architectures are the combination of DenseNet and recurrent units for TAK1 and an ensemble of bi-directional LSTM models for UHL2.
The proposed D3Net exhibited the best performance for \textit{vocals}, \textit{drums} and  \textit{accompaniment} (the summation of \textit{drums}, \textit{bass} and \textit{other}) and performed comparably to the best method for \textit{other}. The average SDR of four instruments is significantly better than all baseline values.
The primally difference between MMDenseLSTM (TAK1) and the proposed method is that MMDenseLSTM incorporates LSTM units to further expand the receptive field, whereas the proposed method uses the multidilated convolution. Comparison of these methods indicate the effectiveness of the multidilated convolution.
On the other hand, GRU dilation 1 \cite{Liu19} consists of dilated convolution and dilated GRU units without a down--up-sampling path. This also highlights the effectiveness of the multiresolution modeling of the multidilation convolution with the dense connection.
For \textit{bass}, approaches that operate in the time domain perform better, as they are capable of recovering the target phase, which is easier in the low frequency range. Among the frequency domain approaches, D3Net performs the best.

We also conducted an ablation study to validate the effectiveness of the multidilated convolution. By replacing the multidilated convolutions with the standard convolutions without dilation, we obtained comparable results as the best performing model in SiSEC2018, TAK1 (MMDenseLSTM). When we replaced the multidilated convolution with the standard dilated convolution, we obtained a decent improvement over the D3Net without dilation even though the aliasing problem arises. However, the proposed multidilated convolution clearly outperforms the standard dilated convolution, showing the importance of handling the aliasing problem in order to incorporate dilation in DenseNet. 

\begin{figure}[t]
  \centering
  \includegraphics[width=\linewidth]{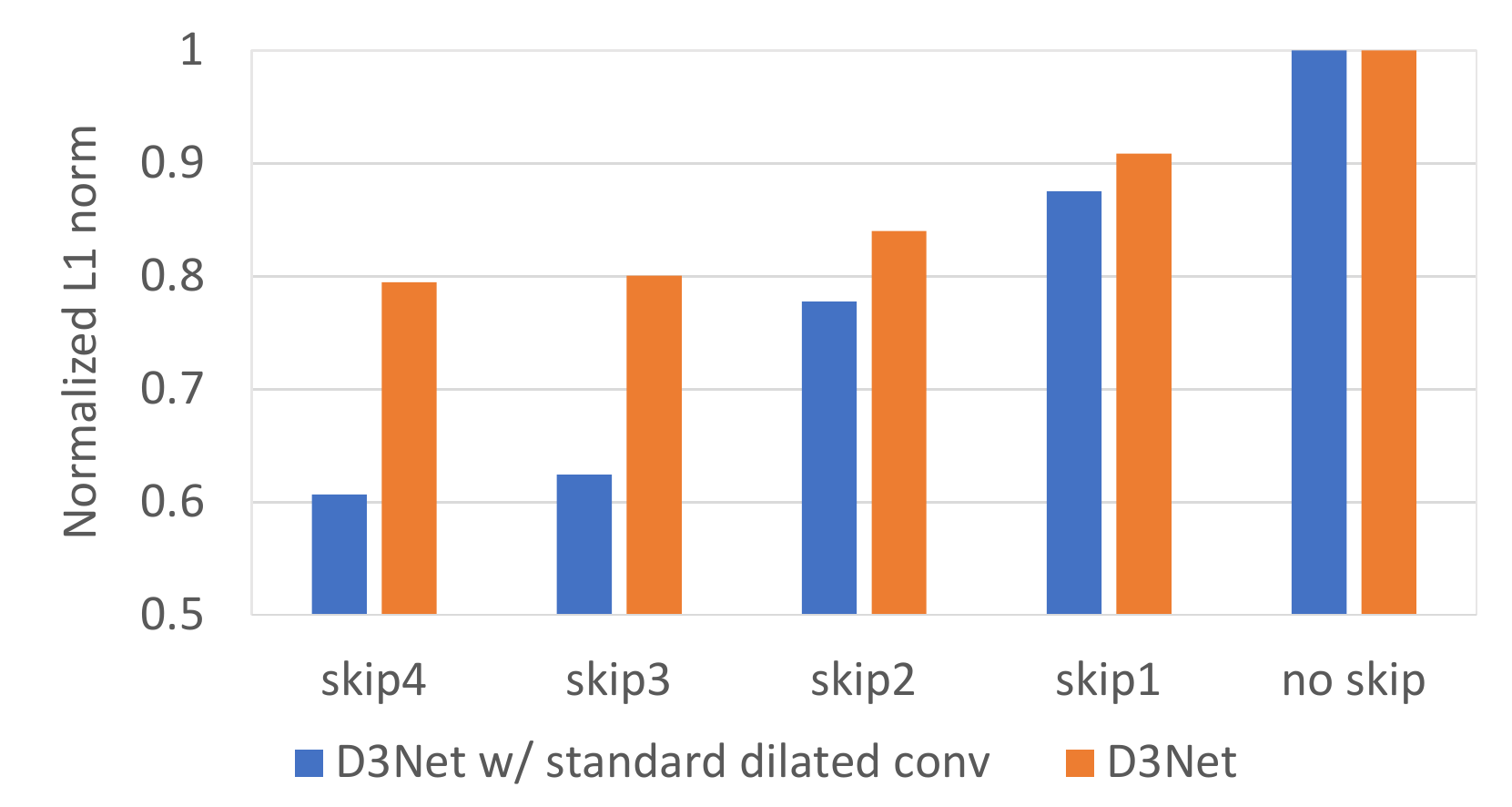}
  \caption{Comparison of normalized L1 norm of weights in the last layer of first d3 block.}
  \label{fig:norm}
\end{figure}

\begin{table*}[t]
\caption{\label{tab:mssex} {\it Comparison of models that use external data for training. SDR values are for MUSDB18 test set. '*' denotes method operating in time domain.}}
\vspace{2mm}
\centering{
\resizebox{0.8\linewidth}{!}{
\begin{tabular}{ c | c | c c c c c c} 
\hline
\multicolumn{1}{c|}{} & & \multicolumn{6}{c}{SDR in dB}\\
Method      & Extra data&	Vocals	&	Drums	& Bass & Other  & Acco.& Avg.\\
\hline\hline
TAK2 {\scriptsize (MMDenseLSTM)} \cite{Takahashi18MMDenseLSTM}\ & 800 songs	&	7.16 & 6.81 & 5.40 & 4.80 & 13.73 & 6.04\\
demucs* \cite{defossez2019demucs}\ & 150 songs	& 7.05 & 7.08 & 6.70 & 4.47 & - &6.33\\
Spleeter \cite{spleeter2020} & 24,097 songs / 79 hours & 6.68 & 6.71 &  5.51 & 4.02 & 12.54 & 5.78\\
TasNet* \cite{Lancaster20,Luo18cTAS} 	& 300 hours & 7.34 & \textbf{7.68} & \textbf{7.04} & 4.04 & 13.76 & 6.52\\
\hline
{\bf D3Net} (proposed)	& - & 7.24	& 7.01 & 5.25 & 4.53 & 13.52 & 6.01\\
{\bf D3Net} (proposed)	& 1,500 songs / 93 hours & \textbf{7.80}	& 7.36 & 6.20 & \textbf{5.37} & {\bf 14.26} & \textbf{6.68}\\
\hline
\end{tabular}
}
}
\end{table*} 
We further investigated the effects of the multidilated convolution by assessing the learned weights.
We calculated the L1 norm of the convolution weights in the last layer of the first block for both the proposed D3Net with the multidilated convolution and the baseline D3Net with the standard dilated convolution. The norm was calculated separately for each skip connection and norm values were normalized by the norm of the path with no skip connection. Fig. \ref{fig:norm} shows that the weights of the skip connection from early layers have smaller norms than later layers. This trend is much more prominent for D3Net with the standard dilated convolution than for D3Net with the proposed multidilated convolution. This results also indicate that applying a dilated convolution to skip connections from early layers without handling the aliasing problem makes it difficult to extract information from them and, therefore, the network assigns a low norm to them.

Finally, we trained D3Net with 1500 extra songs to study how much D3Net can be generalized by using larger dataset. In Table \ref{tab:mssex}, we summarize the SDR values of D3Net and other methods that utilize extra data in addition to MUSDB. Although these methods are not directly comparable since the extra data are different for every methods, we observe that the performance of D3Net is greatly improved with the aid of extra data, and obtain state-of-the-art results on \textit{vocals}, \textit{other}, \textit{accompaniment} and the average SDR.

\section{Conclusion}
We proposed a novel neural network architecture called D3Net. D3Net employs the multidilated convolution with dense skip connections that enables the local and global feature information to be modeled simultaneously within a single layer. Experimental results showed that D3Net achieves state-of-the-art results for the MUSDB18 dataset. The ablation study demonstrated the importance of handling the aliasing problem when we combine DenseNet with the dilated convolution.

\itemsep 30pt 
\bibliographystyle{IEEEbib}
\bibliography{bss,cv}

\end{document}